\documentclass[fleqn,usenatbib]{mnras}

\usepackage{newtxtext,newtxmath}

\usepackage[T1]{fontenc}
\usepackage{ae,aecompl}

\usepackage{graphicx}	
\usepackage{amsmath}	
\usepackage{hyperref}

\title[MAXI~J1535$-$571 multi-$\lambda$ variability]{Fast infrared variability from the black-hole candidate MAXI~J1535$-$571 and tight constraints on the modelling}

\author[F. M. Vincentelli et al.]{F. M. Vincentelli$^{1}$\thanks{E-mail: f.m.vincentelli@soton.ac.uk}, P. Casella$^2$, D. M. Russell$^3$, M. C. Baglio$^3$,  A. Veledina$^{4,5,6}$, \newauthor{T. Maccarone$^7$,J. Malzac$^8$, R. Fender$^9$, K. O'Brien$^{10}$, P. Uttley$^{11}$}
\\
$^{1}$Department of Physics and Astronomy, University of Southampton, SO17 1BJ, UK\\
$^{2}$INAF, Osservatorio Astronomico di Roma 
Via Frascati 33, I-00078 Monteporzio Catone, Italy\\
$^3$Center for Astro, Particle and Planetary Physics, New York University Abu Dhabi, PO Box 129188, Abu Dhabi, UAE\\
$^4$Department of Physics and Astronomy, FI-20014 University of Turku, Finland\\
$^5$Nordita, KTH Royal Institute of Technology and Stockholm University, Roslagstullsbacken 23, SE-10691 Stockholm, Sweden\\
$^6$Space Research Institute of the Russian Academy of Sciences, Profsoyuznaya Str. 84/32, Moscow, 117997, Russia\\
$^7$Texas Tech University, Physics \& Astronomy Department, Box 41051, Lubbock, TX 79409-1051\\
$^8$IRAP, Universite' de Toulouse, CNRS, UPS, CNES, Toulouse, France\\
$^9$Department of Physics, Astrophysics, University of Oxford, Denys Wilkinson Building, Keble Road, Oxford, OX1 3RH, UK\\
$^{10}$Department of Physics, Durham University, South Road, Durham, DH1 3LE, UK\\
$^{11}$Astronomical Institute, Anton Pannekoek, University of Amsterdam, Science Park 904, NL-1098 XH Amsterdam, Netherlands\\
}
\date{Accepted XXX. Received YYY; in original form ZZZ}

\pubyear{2020}
\begin{document}
\label{firstpage}
\pagerange{\pageref{firstpage}--\pageref{lastpage}}
\maketitle

\begin{abstract}
We present the results regarding the analysis of the fast X-ray/infrared (IR) variability of the black-hole transient MAXI~J1535$-$571. The data studied in this work consist of two strictly simultaneous observations performed with {\it XMM-Newton} (X-rays: 0.7$-$10~keV), VLT/HAWK-I ($K_{\rm s}$ band, 2.2~$\mu$m) and VLT/VISIR ($M$ and  $PAH2$\_$2$ bands, 4.85 and 11.88~$\mu$m respectively). The cross-correlation function between the X-ray and near-IR light curves shows a strong asymmetric anti-correlation dip at positive lags. { We detect a near-IR QPO  (2.5 $\sigma$) at $2.07\pm0.09$ Hz simultaneously with an X-ray QPO at approximately the same frequency ($f_0=2.25\pm0.05$).} From the cross-spectral analysis a lag consistent with zero was measured between the two oscillations. 
We also measure a significant correlation between the average near-IR and mid-IR fluxes during the second night, but find no correlation on short timescales.  We discuss these results in terms of the two main scenarios for fast IR variability (hot inflow and jet powered by internal shocks). In both cases, our preliminary modelling suggests the presence of a misalignment between disk and jet.
\end{abstract}




\section{Introduction}

Black hole low mass X-ray binaries (BH LMXB) have been historically studied mainly in X-rays,  where, depending on the accretion regime,  great part of the dissipated gravitational energy is radiated away by either a geometrically-thin, optically-thick accretion disc \citep{shakurasunayev} or a geometrically-thick, optically-thin inflow  \citep{esin1997,poutanen1997}. 
However, multiwavelength studies have shown that bright (non-thermal) emission is also present at lower (from optical-UV down to radio) frequencies \citep[see e.g.:][]{fender2001,corbel2002,hynes2003,gandhi2011}. This low-energy emission is usually interpreted in terms of synchrotron radiation from either a hot magnetised geometrically-thick, optically-thin inflow, or a compact collimated jet \citep{veledina2013,malzac2004,malzac2014}.

In the last 20 years the study of sub-second optical-infrared (O-IR) variable emission from BH LMXBs has improved significantly our understanding of these systems. 
The first sub-second observations in the optical band revealed a complex phenomenology \citep{kanbach2001,malzac2003,malzac2004,hynes2003}. 
In particular, the first Cross-Correlation Function (CCF) of XTE~J118+40 showed an intriguing anti-correlation at negative lag (corresponding to X-ray delays), also known as the ``precognition dip''.  This phenomenon was then confirmed by optical/X-ray observations of Swift~J1753.5$-$0127 \citep{durant2008}. 
Multi-epoch observations of the latter source showed evidence for an evolution of the correlation, from the 1-2 second positive response at the outburst peak to the precognition dip structure in the tail \citep{hynes2009,veledina2017}. Due to the  weak radio emission of this object \citep{soleri2010,tomsick2015}, the described behaviour was associated to the hot inflow. In particular, a detailed modelling of this source showed that the observed CCF could be reproduced by assuming that the optical emission originates from both reprocessed and synchrotron radiation coming from a hot, magnetised inflow, while the X-rays are generated from the Comptonization of the synchrotron radiation \citep{veledina2011}.

Further fast O-IR photometry observations led to the discovery of a 0.1-second O-IR lag with respect to the X-ray variability \citep{gandhi2008,casella2010}. 
This behaviour was attributed to the jet and was well reproduced by the so called \lq\lq internal shock model" \citep{kobayashi1997,spada2001,jamil2010} when linking the shells' velocities to the variations observed in X-rays \citep{malzac2013,malzac2014}. This  demonstrated that fluctuations from the inflow can also be transferred into the outflow and opened the possibility to put new constrains on the jet parameters \citep{casella2010,kalamkar2016,gandhi2017,vincentelli2019}.  Further multi-wavelength observations permitted to deeply characterize the O-IR variability and to study the physical processes that take place inside these outflow.  For example, there is now growing evidence that the O-IR jet emitting region is extended and probably stratified \citep{vincentelli2018,vincentellicasells2019,paice2019}; moreover, \citet[][]{malzac2018} recently showed that Doppler boosting modulation can also lead  to an anti-correlation with X-rays on long timescales. 

Another important feature which gave new insight to the geometry of these systems are the quasi-periodic oscillations (QPO). These oscillations are very common in X-rays and have been attributed to Lense-Thirring precession of the hot inflow \citep{stella1998,ingram2012,motta2015}, although this interpretation is still somewhat matter of debate \citep{ingram_notta2020,marcel2020a,marcel2020b,ma_nat}. 
At lower energies, similar and simultaneous QPOs in the optical band have been explained in terms of synchrotron radiation from a precessing magnetized inflow \citep{veledina_poutanen_ingram_2013,veledina2015}. 
In GX~339$-$4, the first QPO observed in the IR band \citep{kalamkar2016} was found in harmonic relation with an X-ray QPO, similar to its optical counterpart found some years earlier \citep{motch1983}. 
Both precessing hot inflow and the inflow plus jet can qualitatively account for this behaviour, but currently the quantitative agreement has been proved only for the latter alternative, using the internal shocks model \citep{malzac2018}.

MAXI~J1535$-$571 is an X-ray transient discovered by the Monitor of All-sky X-ray Image Gas Slit Camera on the International Space Station \citep{matsuoka} and by the Burst Alert Telescope on board of the \textit{Neil Gehrels Swift Observatory} in September 2017 \citep{negoro2017a,kennea2017}. From the first observations it was noticed that the source displayed strong X-ray variability, including low-frequency QPOs \citep{mereminski2017,gendreau2017}, and very bright radio emission with a flat radio spectrum \citep{russelt2017}. Strong emission in the IR was also observed \citep{dincer2017}. This led to identify the source as a BH LMXB.
During the outburst, despite being heavily absorbed (neutral column density $N_{\rm H}\approx10^{22}$~cm$^{-2}$),  the source reached extremely high flux levels, up to 5 Crab in the 2$-$20~keV band \citep{shidatsu2017}.

An intensive multiwavelength campaign was rapidly coordinated to track the evolution of the source \citep{sivakoff2017}. 
Due to its position in the sky, only few observations could have been made with the Very Large Telescopes (VLTs) before the source { became too low on the horizon.}
In this paper we present the results from two strictly simultaneous near-IR and X-ray observations, aimed at studying the fast multi-wavelength variability of this source. 
We complement the study using also, for the first time,  strictly simultaneous mid-IR observations.
We find that the properties of fast near-IR variability put tight constraints on the current jet and accretion flow models.

\section{Observations}
\begin{figure*}   	
   	\includegraphics[width=0.8\textwidth]{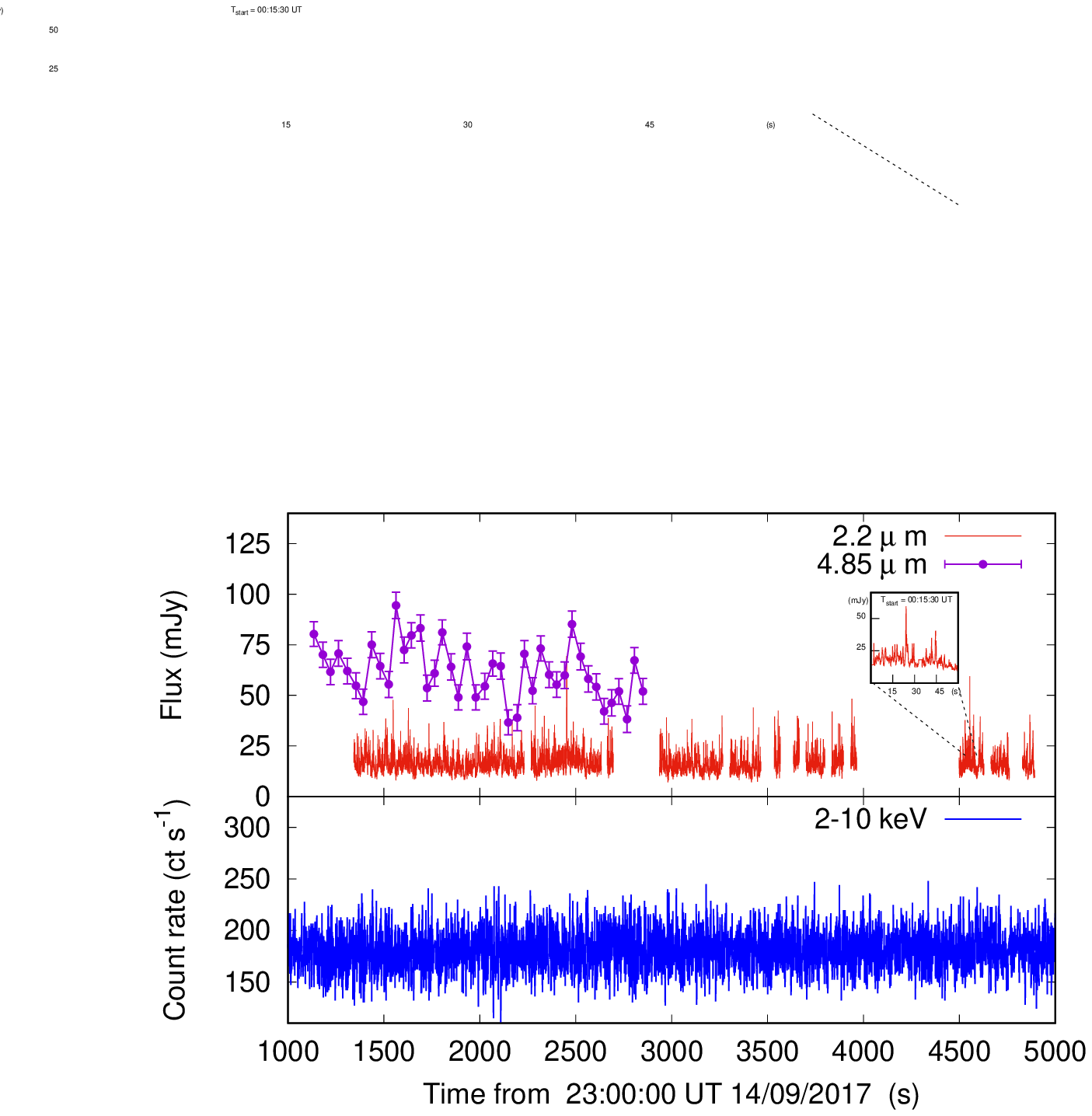}\\
   	  	\includegraphics[width=0.8\textwidth]{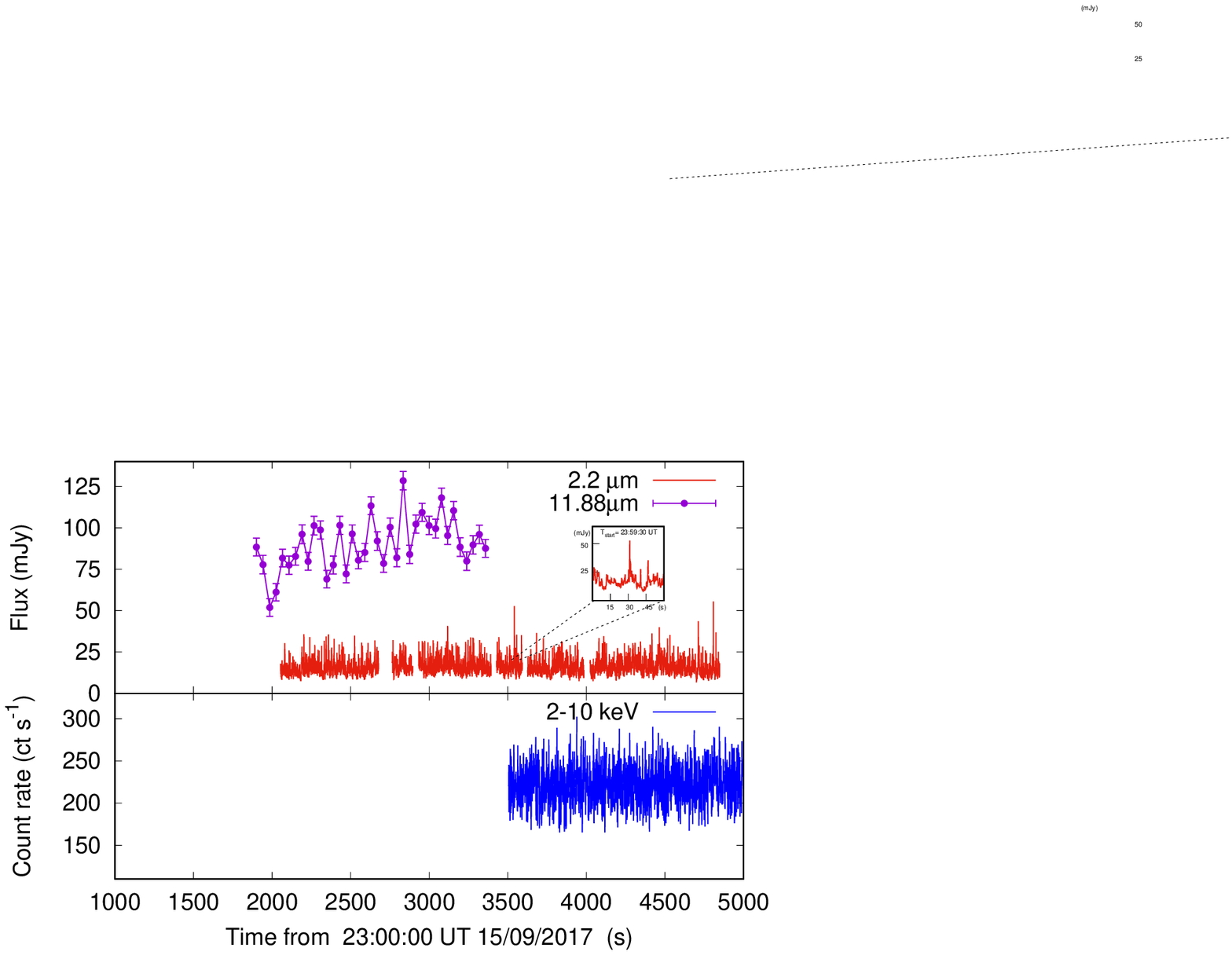}

   	\caption{Lightcurves during X-ray /  near- and mid- IR strictly simultaneous windows for the night of the 14th (top panel) and 15th (bottom panels) of September 2017 (N.B. X-ray observation for the 15th started around midnight, i.e. at $\approx$3600s, (see Table~\ref{tab:obs}). The plot shows the near- and mid-IR (red and purple, respectively) \textit{observed} fluxes (not de-reddened) as a function of time.  Blue curve shows the 2$-$10~keV count rate. While for the near- and mid- IR light curves we used the time resolution of the original data, the X-ray light curve was rebinned with resolution {1s}.	}
   	\label{fig:all_lc}
   \end{figure*}

Strictly simultaneous multi-wavelength observations were taken during the night of the 14th and the 15th of September 2017 (MJD 58010--58011), while the source was in its hard-intermediate state \citep{baglio2018,bhargava2019,russell2018}. Near- and mid-IR data were collected with HAWK-I and VLT Imager and Spectrometer for the mid-infrared (VISIR) respectively, both mounted at the ESO VLTs; X-rays data were collected with the EPIC-pn camera on board the ESA satellite {\it XMM-Newton} (see Fig. \ref{fig:all_lc}).
The epochs of the observations are reported in Table~\ref{tab:obs}.

    \begin{table*}
\caption{Summary of the multi-wavelength campaign on the BH LMXB MAXI~J1535$-$571 on the 14th and 15th of September 2017.}
\label{tab:obs}

\begin{tabular}{ccccc}
     \hline

Night of the Start Time & Telescope / Instrument & Energy Band        & Start Time  (UT) & End Time (UT)  \\
     \hline

      &                        &                    &             &                   \\
      & {\it XMM-Newton}/Epic-PN  & 0.5$-$10 keV         & 16:47:04         & 01:08:44       \\ 
 14/09     & UT-4/HAWK-I          & $K_{\rm s}$ (2.2 $\mu$m) & 23:26:21         & 00:21:08       \\
      & UT-3/VISIR           & $M$ (4.85~$\mu$m)    & 23:19:22         & 23:46:20       \\
      &                        &                    &                  &                \\
\hline
      &                        &                    &             &                   \\
      & {\it XMM-Newton}/Epic-PN   & 0.5$-$10 keV         & 23:59:09         & 04:20:02     \\ 
15/09       & UT-4/HAWK-I          & $K_{\rm s}$ (2.2 $\mu$m) & 23:34:15         & 00:19:25       \\
      & UT-3/VISIR           & $PAH2$\_$2$ (11.88 $\mu$m)    & 23:29:54         & 23:53:26       \\
      &                        &                    &                  &                \\
\hline 
      
\end{tabular}
\end{table*}

\subsection{X-ray data. \textit{XMM-Newton}}

X-ray data were collected with the EPIC-pn camera. Observations were taken in \textit{Burst mode} (OBS ID: 0795712001 and 0795712101). The data reduction was carried out using the XMM Science Analysis System (SAS v15). In particular, the source counts were extracted in the range RAWX: 28$-$48 (corresponding to an angular size of $\approx$86~arcsec along RAWX). Due to the high absorption of the source, X-ray events were extracted only between 2 and 10 keV. The mean count rate of the two observations was found to be very similar: $234\pm1$~cts~s$^{-1}$ for the first night, and $226\pm1$~cts~s$^{-1}$ for the second. Data were then  barycentered with the \textit{barycen} software and binned in a light curve of 5.7 ms.

\subsection{Near-IR data. VLT/HAWK-I}


Near-IR sub-second data were collected with HAWK-I \citep{pirard}  mounted at the VLT UT-4/Yepun under Program: 099.D-0068(A).
1-hour long observations were taken in the $K_{\rm s}$ band (effective wavelength 2.2~$\mu$m) with the time resolution of 0.125~s. Images were stacked in ``data-cubes'' of 250 frames, separated by a gap of $\approx$3~s.
A fraction of the cubes was affected by frame-losses.
For cross-correlation analysis cubes with frame losses were therefore discarded in order to avoid to introduce spurious lags. During the observation, (especially in the second part of the first night), the target ended up few pixels from the edge of the detector, introducing spurious features in the lightcurve. To be conservative we decided not to include those cubes in the analysis (hence the large gaps seen in Fig. \ref{fig:all_lc}).

The telescope was pointed to 15:35:41.6, $-$57:11:22.9 (RA, DEC) at an angle of 70\degr\,in order to have the target on the lower-left quadrant (Q1) together with the reference star with a $K_{\rm s}$ magnitude of 11.55$\pm$0.02. 
The data reduction was performed using the ULTRACAM data reduction software tools \citep{dhilon2007}. 
Parameters for the reduction were derived from the bright reference star point spread function and position.  
To account for seeing and transmission variations, the ratio between the source and the bright reference star count rate was used. 
Individual power spectral densities (PSDs) of the non-variable reference and comparison star showed that the variability induced by the seeing is extremely low, as compared to the intrinsic variability of the target, and was successfully removed by taking the ratio between the target and the bright reference star.
Using such extraction, we found an average $K_{\rm s}$ magnitude of 11.52$\pm$0.01 (non-dereddened flux of $\approx$16~mJy) for both nights.  This in good agreement with the near-IR data in \citet{baglio2018} taken on the same nights by REM. The time of each frame was then put in the Dynamical Barycentric Time system.

\subsection{Mid-IR data. VLT/VISIR}

Mid-IR observations were obtained with VISIR \citep{lagage04} mounted at the VLT-UT3 Melipal under Program 099.D-0884(A). The instrument was set in small-field imaging mode, with a pixel scale of 45$\, \rm mas\, \rm pixel^{-1}$, and the $M$ and $PAH2$\_$2$ filters were used on September 14th and 15th, respectively (covering the $4.85-11.88\,\mu$m wavelength range; see Table~\ref{tab:obs}). Every observation consisted of 1000s of time on source, characterized by 44 nodding cycles. In total, considering chopping and nodding between sky and source, the total exposure time was 1800--1900~s per observation.

Reduction of data was performed using the VISIR pipeline, available in $gasgano$\footnote{\url{ https://www.eso.org/sci/software/gasgano.html}}. Sensitivities were estimated thanks to the observation of two standard stars (HD161096 and HD163376) on the same night and telescope configuration. After recombination of the chop/nod cycle raw images, aperture photometry was performed by means of an aperture that was large enough to avoid that possible seeing variations could affect the portion of the flux falling in the aperture. 
We note that the target of the observations was bright enough to allow a detection in individual observations. The VISIR pipeline requires that observations are combined in groups of multiples of two. In order to achieve a flux from each observation for this work, we therefore had to combine each image with the first observation on each night, then take into account the flux of this first observation to obtain the light curve. Using this method we were able to sample the data twice as fast with respect to  \citet[][]{baglio2018}. In particular, each time bin has  27.8 s exposure, with a varying gap of $\approx$ 10 seconds between the bins due to the chop/nod cycle and read-out. Also in this case, the resulting lightcurve was then put in Dynamical Barycentric Time system.

No clear variability of the background was detected during the entire VISIR observations (see also \citealt{baglio2018} for details). A variation of $(7-8)\%$ in the ADU/flux conversion factor over different dates has instead been observed. 
This variability can however be caused by the different weather and sky conditions for different dates, as well as different air masses. Therefore, any possible mid-infrared variability of MAXI~J1535$-$571 can be safely considered as intrinsic to the source.

\section{Data Analysis}

\subsection{X-ray and near-IR power spectral densities}

The properties of the X-ray and near-IR variability were studied through Fourier power spectral analysis.
The X-ray and near-IR light curves were divided in 16384-bin and 512-bin long segments, respectively. 
We computed the power spectral density (PSD) of each segment in squared root-mean-square (rms)  normalization \citep[see e.g.][]{belloni1990,miyamoto1991,vaughan2003}  and then applied a logarithmic binning factor of 1.05 to the resulting average. The gaps between the cubes in the near-IR data were filled with simulated data, following the procedure described in \citet{kalamkar2016}. For the near-IR case, in order to reach lower frequencies without affecting the statistics of the higher frequencies, we also computed a PSD with the light curve rebinned to 10 s (much longer than the $\approx$3~s gaps) using 16 bins per segment. 
   \begin{figure*}
\centering
\includegraphics[width=\columnwidth]{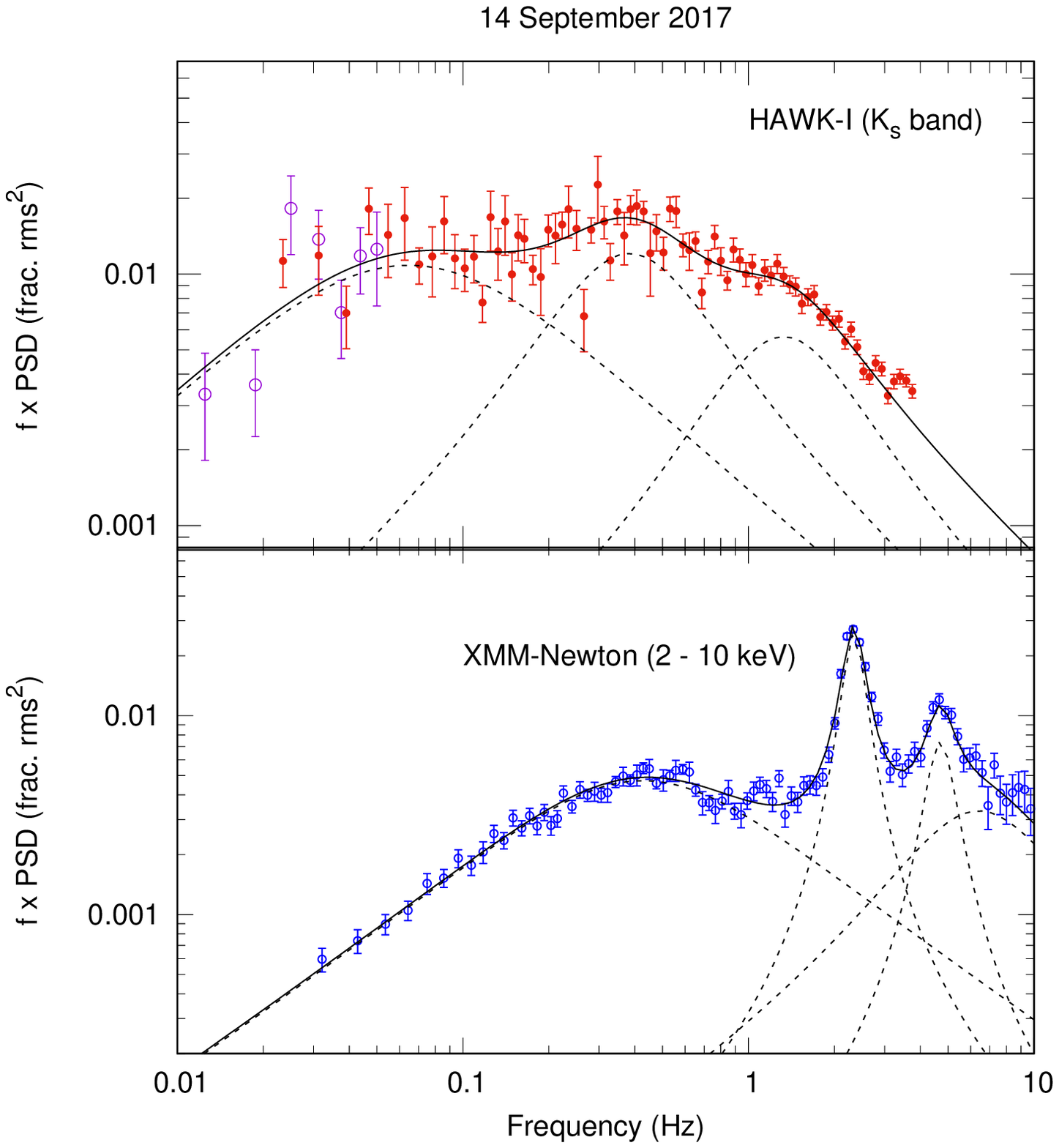}\hspace{0.5cm}
\includegraphics[width=\columnwidth]{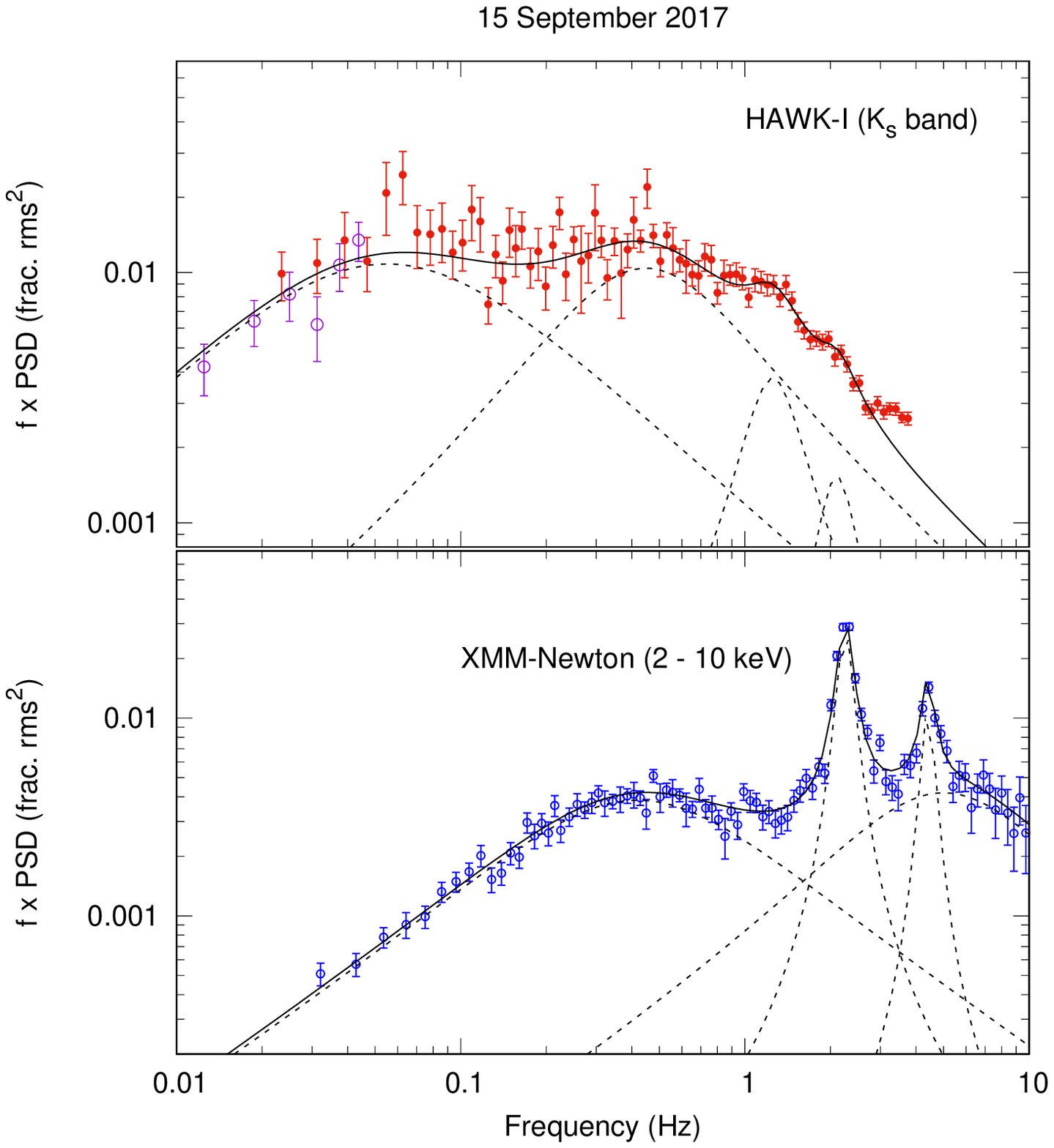}
\caption{PSDs computed from the HAWK-I and {\it XMM-Newton} light curves for the night of 2017 September 14th and 15th. The models fitted to the PSDs and their individual components are shown with black and dashed lines, respectively. Data above 3~Hz are not used for fitting purposes due to clear aliasing contribution. Empty purple points represent the near-IR PSD computed with a 10~s light curve and 16 bins per segment.
}

\label{figure:psd_1535}
\end{figure*}

   \begin{table*}
   
    \caption{Parameters of the fit to the PSD with multiple Lorentzian components, $L(f)=A/ [(f-f_0)^2+(\Delta/2)^2]$.
    We additionally define Q=$f_0/\Delta$.
    The fractional rms of each Lorentzian was obtained from the squared of the integration over the whole frequency range .}
    \label{tab:fits_pds}
    \begin{tabular}{ccccccccc}
     \hline

Date & Band &  Comp. & $f_0$ & $\Delta$ & A  & rms & Q & $\chi^2/$d.o.f.\\
  & &    & (Hz) & (Hz) &  ($10^{-3}$~Hz) &  & & \\
     \hline
   \newline

14/09 & X-rays &  1 & 0.12 $\pm$ 0.02  & 0.82 $\pm$0.02 & 13 $\pm$ 1 &  0.356   $\pm$  0.001  & 0.14 $\pm$ 0.02 & 140/102\\
  & &   2  & 2.33 $\pm$ 0.01 & 0.46 $\pm$ 0.02 & 7.9 $\pm$ 0.3 & 0.089   $\pm$  0.004 & 2.8 $\pm$ 0.07& \\
  & &   3  & 4.66 $\pm$ 0.05 & 1.2 $\pm$ 0.3   & 3 $\pm$ 1 &  0.05   $\pm$  0.01 & 4 $\pm$ 1   \\
  & &   4  & 5 $\pm$ 1       & 9 $\pm$ 1       & 5 $\pm$ 2 &  0.07   $\pm$  0.02 & 0.6 $\pm$ 0.1   \\

   \hline
   \newline
 15/09 & X-rays & 1 & 0.15 $\pm$ 0.03    & 0.78 $\pm$0.06 & 10 $\pm$ 1 &  0.08   $\pm$  0.01  & 0.19 $\pm$ 0.04 & 101/102\\
  & &   2 & 2.25 $\pm$ 0.05 & 0.31 $\pm$ 0.02 & 6.4 $\pm$ 0.3 & 0.08   $\pm$  0.01 & 7.3 $\pm$ 0.5& \\
  & &   3  & 4.4 $\pm$ 0.03 & 0.5 $\pm$ 0.1 & 1.8 $\pm$ 0.2 &  0.05   $\pm$  0.01 & 9 $\pm$ 2   \\
  & &   4  & 3 $\pm$ 1 & 9 $\pm$ 2 & 9 $\pm$ 2 &  0.08   $\pm$  0.02 & 0.3 $\pm$ 0.1   \\

   \hline
   \hline
   \newline

 14/09 & near-IR &   1 & 0 (fixed)      & 0.13 $\pm$0.02  & 3.4 $\pm$ 3 &  0.13   $\pm$  0.02  & - & 83/80\\
       &         &   2 & 0.3 $\pm$ 0.04 & 0.51 $\pm$ 0.08 & 22 $\pm$ 5 & 0.14   $\pm$  0.03 & 2.2 $\pm$ 0.1& \\
       &         &   3 & 1.1 $\pm$ 0.2  & 1.6 $\pm$ 0.2   & 9 $\pm$ 2 &  0.08   $\pm$  0.02 & 0.7 $\pm$ 0.2   \\

   \hline
   \newline

 15/09 & near-IR & 1 & 0 (fixed)        & 0.11 $\pm$0.01  & 34 $\pm$ 3    &  0.13   $\pm$  0.02  & - & 86/79 \\
       &         & 2 & 0.28 $\pm$ 0.076 & 0.69 $\pm$ 0.12 & 22 $\pm$ 4    & 0.29   $\pm$  0.02 & 0.4 $\pm$ 0.1& \\
       &         & 3  & 1.2 $\pm$ 0.1   & 0.6 $\pm$ 0.2   & 3 $\pm$ 1     &  0.05   $\pm$  0.01 & 1.5 $\pm$ 0.6   \\
       &         & 4 & 2.07 $\pm$ 0.09  & 0.77 $\pm$ 0.25 & 0.8 $\pm$ 0.3 & 0.03   $\pm$  0.01 & 3 $\pm$ 1& \\

     \hline
    \end{tabular}
   \end{table*}

Fig.~\ref{figure:psd_1535} shows the PSDs for both epochs in X-rays and the in near-IR.
The X-ray PSD does not change significantly between the two nights.
In order to quantify the overall variability, a fit with multiple Lorentzian components was performed \citep{belloni2002}. 
Fig.~\ref{figure:psd_1535} (bottom panels)  shows how the broadband noise can be well fitted by two Lorentzians centered at  $\approx$0.5~Hz and  $\approx$5~Hz, respectively, while a type-C QPO \citep{casella2005,motta2015} and its second harmonic are clearly detected at $\approx$2.25~Hz and 4.5~Hz. 
These values are consistent with the values measured from quasi-simultaneous {\it Astrosat} X-ray timing observations \citep{bhargava2019,sreehari2019}.
Results from the fit are reported in Table~\ref{tab:fits_pds}.

The near-IR PSDs show clear differences with respect to the X-rays:  the low-frequency variability appears to dominate the IR PSDs, with a clear break at around 1~Hz, while the X-ray PSDs are dominated by variability in the 0.1$-$10~Hz frequency range.
Similarly to the X-rays, we modelled the near-IR PSDs with a number of Lorentzian components. The broadband noise, which in this case extends up to the Nyquist frequency, shows an excess at the highest frequencies, suggesting the presence of aliasing.
Therefore, the fit was performed excluding frequencies above 3~Hz. 
During the first night the broadband noise was well described with 3 components, centred at 0 Hz (fixed), $\approx$0.3~Hz and $\approx$1~Hz, the latter one identifying the break visible in the PSD.
During the second night, the centroid frequencies of the three components remain the same, although the high-frequency component narrowed significantly, and an additional fourth, even narrower component was marginally (at $\approx$2.5$\sigma$ level) required by the fit. 
The centroid frequency of this additional component is consistent with that of the QPO detected in the simultaneous X-ray light curve. {In order to confirm this hypothesis we fixed the frequency and width of this component to the one measured in the X-rays, leaving only the normalization as a free parameter. The amplitude of the QPO in this case was found significant at a 3 $\sigma$ level, strengthening the evidence for a QPO in the infrared band. }

\subsection{X-ray/near-IR cross-correlation analysis}\label{sec:xir}

\begin{figure}
\centering
\includegraphics[width=\columnwidth]{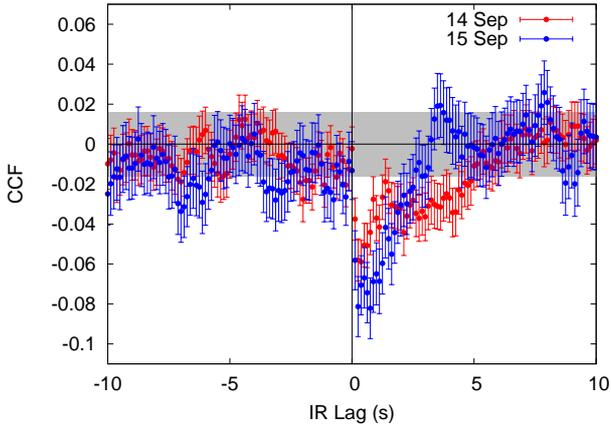}
\caption{Near-IR/X-ray CCF computed using the {\it XMM-Newton}/Epic-PN (2$-$10~keV) and HAWK-I $K_{\rm s}$ light curves for the two nights of observations (September 14 and 15, red and blue, respectively).  The 1$\sigma$ confidence level is shown in grey. An anti-correlation dip at positive lags is clearly detected in both nights.}
\label{fig:ccf_nearIR}
\end{figure}

The correlated variability between the X-ray and near-IR light curves was investigated using both time- and frequency-domain techniques. In order not to distort the results we used only the original light curves (i.e. without filling the near-IR gaps) and excluding cubes affected by frame-losses.
We computed the CCF \citep{edelson1988} for all the simultaneous observations using the procedure described in \citet{gandhi2010}. 
The computed CCFs in the two nights present a strong asymmetric dip at positive lags (positive lags correspond to near-IR lagging behind the X-rays), with a $\lesssim$1-second sharp drop and a $\sim$5-second shallower rise (Fig.~\ref{fig:ccf_nearIR}).

We quantified the possible noise contribution to the observed CCF by simulating $N=10^3$ couples of uncorrelated light curves with the same power-spectral properties of our dataset. The resulting distribution of  of the $N$ uncorrelated CCFs has a standard deviation of 0.016 (grey-shaded area in Fig.~\ref{fig:ccf_nearIR}). The anti-correlation dip appears statistically significant, while the peaks observed at lags larger than $\pm$5~s -- as well as those at negative lags -- are consistent with noise. 

\begin{figure}
\centering
\includegraphics[width=\columnwidth]{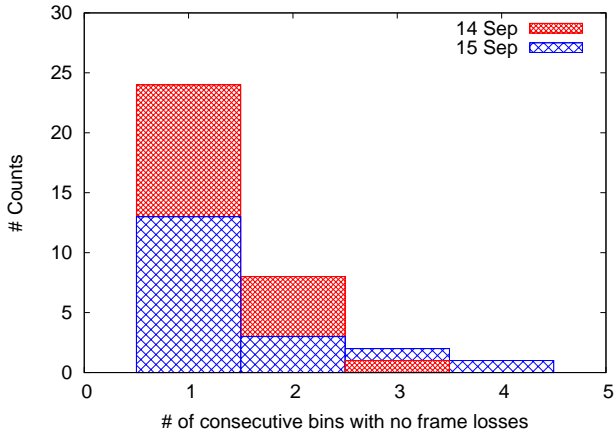}
\caption{Histogram of the number of consecutive data-cubes with no frame losses. For both nights, the majority of the segments of the cleaned light curve consisted of a single data-cube. This means that the distance between the segments was (almost always) too large to be filled with standard methods \citep[see also:][]{kalamkar2016}.}
\label{fig:hist}
\end{figure}

\begin{figure}
   	\includegraphics[width=\columnwidth]{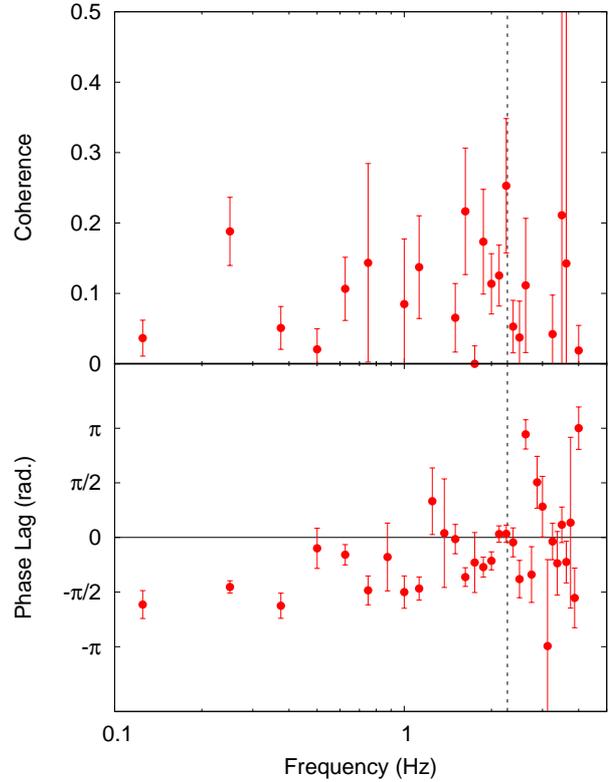}
   	\caption{Intrinsic coherence (top panel) and phase lags (bottom panel) computed between near-IR and X-ray light curves, combining data from both nights. 64-bin segments were used. Vertical dashed line shows the position of the QPO.}
   	\label{fig:coh}
   \end{figure}

To investigate further such correlated variability, Fourier cross-spectral analysis was applied. {In particular, intrinsic coherence and lags were computed using the procedure described in \citet{uttley2014} \citep[see also][ and references therein]{vaughannowak1997,ingram2019}.}
The presence of frames losses forced us to discard a significant fraction of the data-cubes in the cross-correlation analysis. The remaining ``cleaned'' light curve presented many interrupted segments, with gaps much longer than those seen between two consecutive data-cubes (see Fig.~\ref{fig:hist}). 
In order to compute the cross spectral densities, we used 64-bin long segments, which allowed us to explore frequencies down to 0.125~Hz. In order to probe longer timescales, one would have to either deal with much lower statistics (because of the lower number of sufficiently long intervals) or fill the gaps between intervals. In the latter case, even though the resulting statistics can be formally high, the resulting cross spectra would appear significantly distorted, as confirmed by simulations.
However, given the width of the dip in the CCF, frequencies in the $\approx0.1-0.25$~Hz range are still expected to contribute to the observed anti-correlation. Therefore our choice is suitable for our purpose. Due to the low statistics, and given the similar timing properties during the two observations, cross-spectral analysis was performed combining the data from both nights together. We also checked the data for the individual nights but no significant variation from the described results was found. The results are shown in Fig. \ref{fig:coh}.

The intrinsic coherence of the two datasets is very low for almost all the probed frequencies. Nevertheless the phase lags, though scattered, seem to show a clear trend. At low frequencies there is evidence for a (nearly) constant phase lag at $\sim -\pi/2$, while at frequencies higher than the QPO the phase lags are unconstrained, thus oscillate between $-\pi$ and $\pi$.
 A somewhat higher coherence of $0.25\pm0.1$ (1$\sigma$ error) is measured over the frequency range where the QPO is detected in the X-ray light curve.
In this range we also find the phase lag consistent with zero  ($\approx\pi$/6 rad. 3 $\sigma$ upper limit).  {We tested the significance of this variation by integrating the phase lags and coherence within the QPO frequencies (2.05-2.3 Hz) and in an equally large bin, centered just before the QPO (1.7-2.05 Hz). The phase lags pass from  -0.8 $\pm$ 0.2 rad.  outside the QPO,  to 0.05 $\pm$0.1 rad. (1 $\sigma$); the coherence, instead, increases from 0.08 $\pm$ 0.02 to 0.12 $\pm$ 0.02  (1 $\sigma$).  This clearly shows the presence of an additional correlated component associated with the QPOs observed in both X-rays and IR at $\approx$ 2.1 Hz. }

\subsection{Correlations with mid-IR}

The variability properties of the mid-IR light curve have already analysed by \citet{baglio2018}. During the first night (4.85~$\mu$m) the source showed a flux of 62.4~mJy and a fractional rms of 17.2$\pm$ 8.4\%; during the second one (11.88~$\mu$m) a flux of 90.2~mJy and a 14.9$\pm$ 4.8 \% rms were measured. Here we focus on the correlated variability between mid-IR and the near-IR/X-ray light curves. While the correlation with the near-IR band could be computed for both nights,  strictly simultaneous X-ray/mid-IR coverage was obtained only during the first night.

\begin{figure*}
   	\includegraphics[width=\columnwidth]{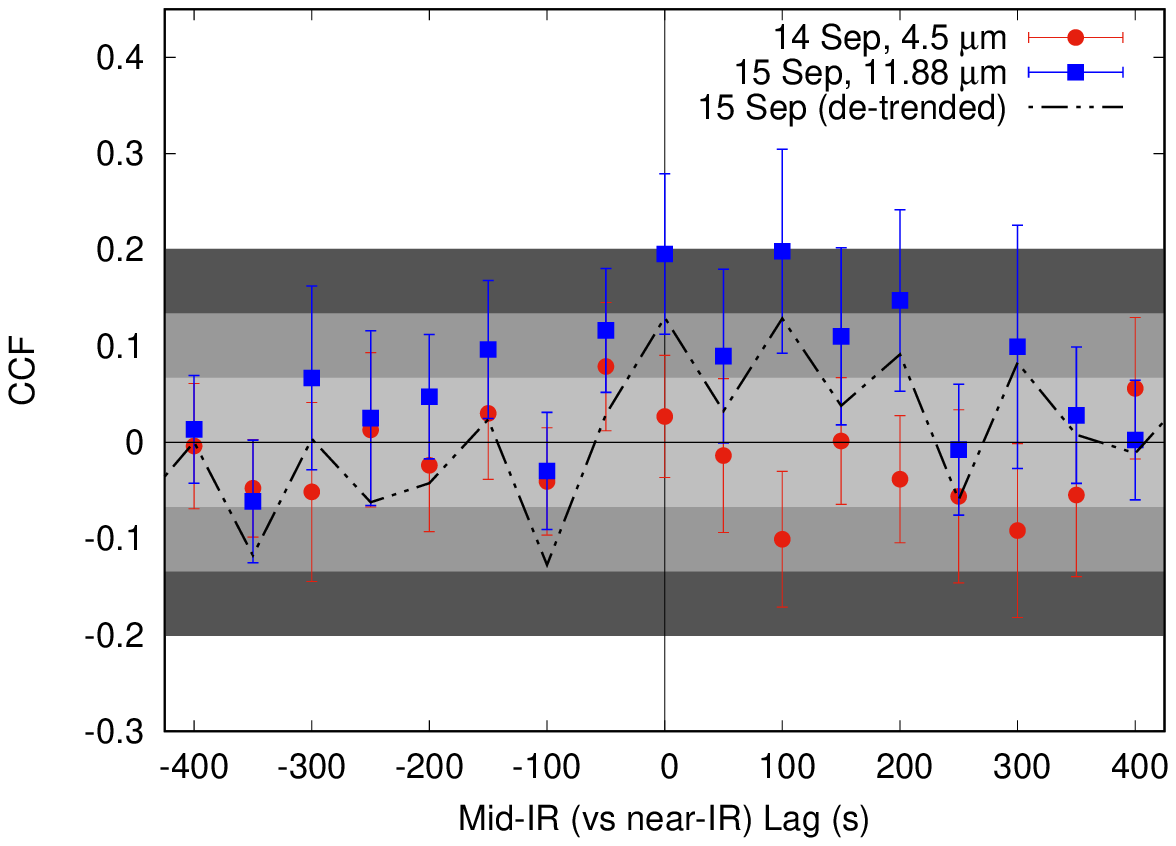}
   	\includegraphics[width=\columnwidth]{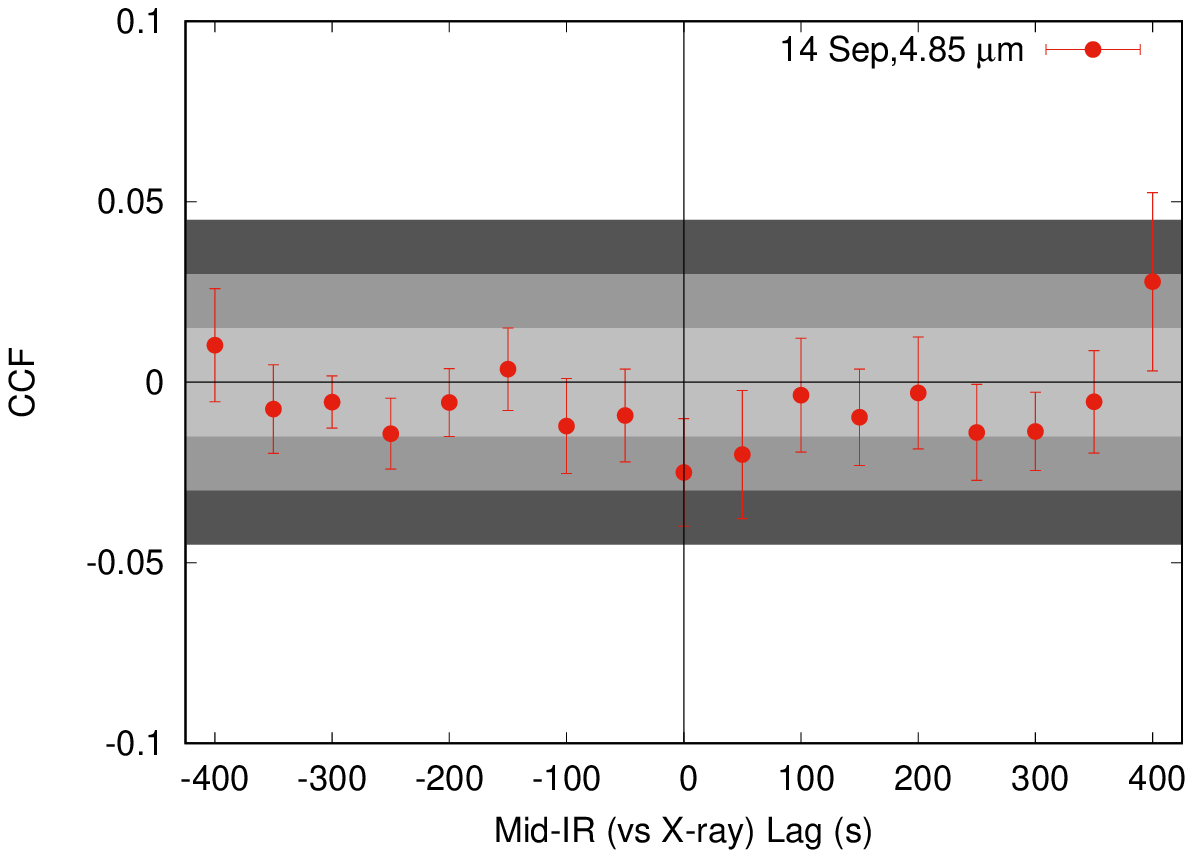}
   	\caption{Mid-IR/near-IR (left) and mid-IR/X-ray (right) CCFs. 1-, 2- and 3$\sigma$ confidence levels are plotted with different grey shades. Dotted-dashed line in the left panel shows the CCF for the second night after the linear de-tremding was applied.}
   	\label{fig:ccfs_mid}
   \end{figure*}

Similarly to the near-IR/X-ray case, we computed the CCF using  50~s time bins.  Errors were estimated following the procedure described in \citet{edelson1988} and \citet{gandhi2010}. We then evaluated the noise contribution by applying the same method described in Sect.~\ref{sec:xir}. 
The CCFs are plotted in Fig.~\ref{fig:ccfs_mid}, with grey shades representing the 1-, 2- and 3$\sigma$ confidence levels. 
For the first night, we did not find any significant correlation between X-ray and mid-IR nor between near-IR and mid-IR light curves. A visual inspection of the light curves and the PSDs suggests that the reason behind this non-detection is the lack of variability for Fourier frequencies lower than $\approx$0.03 Hz (30 s timescales).

\begin{figure}
   	\includegraphics[width=\columnwidth]{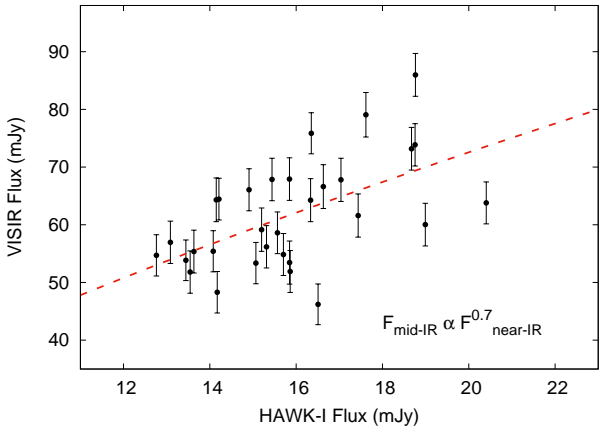}
   	\caption{Mid-IR vs near-IR correlation diagram for the second night. As for Fig. \ref{fig:all_lc} the fluxes are not de-reddened. A correlation of 3 $\sigma$ is detected.}
   	\label{fig:flux-flux}
   \end{figure}

The near-/mid-IR CCF of the second night instead presents two peaks at $\approx 3\sigma$-level at 0~s and 100~s lags (mid-IR lags behind near-IR). 
We note, however, that during the second night, the mid-IR light curve shows a slowly increasing long-term trend, which could affect the CCF. We therefore computed the CCF after removing a linear trend to the light curve (see Fig.~\ref{fig:ccfs_mid}). The correlation in the second case is below 2 $\sigma$ (simulations showed that the confidence levels did not change significantly after the de-trending). This shows that the observed correlation is due to the long-term trend, and not originating from the fast timescales variations. In order to visualize this better, we also plotted the mid-IR vs near-IR correlation diagram (Fig.~\ref{fig:flux-flux}), averaging the HAWK-I values within the 27.8s VISIR exposures. The Pearson correlation coefficient between the two series is $\rho$=0.56, which according to a simple t-distribution  is significant at a  $\approx$3 $\sigma$ level. We also quantified the relation between the near and the mid-IR variations by fitting as a power law $F_{\rm mid-IR}\propto F^\beta_{\rm near-IR}$, finding a slope of $\beta$=0.7$\pm$0.2.

\section{Discussion}

\subsection{Power spectrum and QPO}

We observe strongly variable near-IR emission from the BH LMXB MAXI~J1535$-$571 in its intermediate state. 
The broadband variability extends down to sub-second timescales, albeit with a clear break in the PSD at $\approx$1~Hz. 
Similar behaviour has been reported already in a few BH LMXBs:  \citep{motch1983,casella2010,gandhi2010,kalamkar2016,veledina2017,vincentelli2018}. 
Superimposed on the strong broadband noise, a $\sim2.1$~Hz QPO is simultaneously detected both in the X-rays and (marginally) in the near-IR PSDs, which we identify as a type-C QPO. {It is interesting to notice that the Q factor of the (X-ray) main QPO and its harmonic are consistent within the errors. This has already been observed in other sources \citep[][]{rao,ratti2012} and indicates that the origin of the quasi-periodicity is due to the fluctuations in the frequency rather than the amplitude. This was found to be consistent with the predictions of Lense-Thirring precession of the accretion flow \citep[][see however also \citealp{marcel2020b}]{ingram2009,ingram2012}. }

O-IR type-C QPOs have been reported in a number of BH LMXBs so far \citep{motch1983,gandhi2010,veledina2015}. In particular, GX 339-4 showed IR QPOs in harmonic relation with the X-ray one \citep{kalamkar2016}, or without a clear equivalent in the X-ray PDS \citep[i.e. X-ray QPO marginally  or not detected][]{vincentelli2019}. {Despite being weaker compared to GX 339-4, \citep[3$\pm 1$\% vs 6$\pm 2$\%, see Tab. 2 and Tab. 1. in][]{kalamkar2016},  our IR QPO  is the fastest detected to date (by almost a factor of ten); in addition,  \citep[differently from][]{malzac2018} we also find a zero phase lag. Interestingly,  optical/X-ray lags  consistent with 0 at the QPO frequency have  also been reported from the BH LMXB Swift~J1753.0$-$0127 \citep{veledina2015}, which might suggest a similar physical scenario.}

The origin of similar fast-variable O-IR emission is still matter of debate \citep*{uttleycasella2014,poutanen2014}. 
The two most promising scenarios invoke synchrotron radiation either from a magnetised hot accretion flow \citep{veledina2013}, or from a collimated jet \citep{malzac2014}. 
Using the obtained results of the simultaneous near-IR and X-ray QPOs, we can put constraints on both models.

In the hot flow scenario, the O-IR QPOs are related to the Lense-Thirring (LT) solid-body precession of the hot medium \citep{veledina_poutanen_ingram_2013}. {This model has been mainly tested with QPOs at much lower frequencies \citep[$\approx10^{-3}$-$10^{-2}$ Hz; ][]{veledina2017}, however, as long as the oscillation is also detected in X-rays, there are no principle restrictions on the QPO frequency.} If we assume that the observed near-IR frequency $\nu=1.4\times10^{14}$~Hz is the frequency at which the synchrotron spectrum cuts off (towards longer wavelengths, the so-called turnover frequency, \citealt{wardzinskizdziarski2001}, see also eq.~3 in \citealt{veledina2013}), it is possible to set a limit to the size of the partially-absorbed synchrotron-emitting region.
To comply with the required near-IR frequency, we obtain constraints on the electron Thomson optical depth ($\tau$) of the power-law electrons in the hot flow $\sim10^{-3}-10^{-2}$, and the magnetic field in the medium $\sim10^4-10^5$G, which can be achieved at $R/R_{\rm g}\sim30-50$ (for the black hole mass of 10 solar masses).
This estimate is to be compared to the outer radius of the precessing accretion flow, as inferred from the Fourier frequency of the QPO, $f=2.1$~Hz.
Using eq.~(43) of \citet{fragile2007} for the QPO frequency and substituting the estimate for the inner disc radius from \citet{lubow2002} \citep[see also eq.~2][]{ingram2009}, we obtain that the required radii can be achieved for the spin parameter $a\gtrsim0.8$, radial power-law dependence of the disc surface density $\Sigma\propto R^{-0.5}$ (typical for advection-dominated accretion flows, \citealt{narayan1994}) and $H/R\sim1$. 
Hence, in order to explain the simultaneous near-IR and X-ray QPO within the hot flow model, we require high spin and high disc aspect ratio.
The obtained zero phase lag between the X-ray and near-IR QPO can be explained by the solid-body precession of the hot flow, if the orbital inclination is $\gtrsim50$\degr\, \citep{veledina_poutanen_ingram_2013}.
The high spin and orbital inclination seems to be consistent with recent X-ray spectral measurements \citep{miller2018,xu2018,sreehari2019}.

The alternative scenario which is usually invoked for O-IR QPOs is the LT precession of a relativistic jet together with the X-ray emitting hot inflow.
Recent 3D GRMHD simulations have shown that the jet can precess along with the hot inflow \citep{liska2018}. 
Moreover, \citet{malzac2018} showed that the amplitude and the width of the IR QPO observed in GX~339$-$4  during its hard-intermediate state in 2010 \citep{kalamkar2016} could be reproduced by the precession of a jet, by assuming a partial breaking of the coherence of its oscillations, as also expected during LT precession. 
For instance, if $\Gamma \beta \approx0.2-0.5$ (where $\Gamma$ is the jet Lorentz factor and $\displaystyle \beta=\sqrt{1-\Gamma^{-2}}$ is the dimensionless jet velocity) and a precession angle of $\approx5$\degr\, would produce the required 3\% rms observed in the IR. 

The observed zero phase lag then imposes constraints on the distance between the X-ray and near-IR emission sites. This can be used to constrain the maximum inclination of the jet, and to set a lower limit to its speed ($\beta$). By taking into account projection effects, the distance between the two emitting regions will be  $R = \beta c \Delta t/(1-\beta \cos i)$, where $\cos i$ is the cosine of jet inclination angle with respect to the line of sight, and $\Delta t$ is the near-IR/X-ray time lag. By integrating the cross-spectrum on the frequencies where the phase lag is consistent with 0, we obtained a $3\sigma$ upper limit to the time lag of 0.04 s.
 
Simulations from the internal shocks model have shown that the peak of the near-IR emission region (R) can span, depending on the jet physical conditions, a range of values from  $\approx 5\times10^{3} R_{\rm g}$ to  $\approx 5\times10^{5} R_{\rm g}$ \citep{malzac2014,malzac2018}. Given also the limits to the jet first acceleration zone imposed by \citet{russell2020}, we set a conservative value for R of $5\times10^{3} R_{\rm g}$, which leads to a required jet Lorentz factor of $\Gamma\gtrsim$1.7 and a jet inclination $i\lesssim 40$\degr.
It is interesting to notice that this is  in good agreement with  radio observations of MAXI~J1535$-$571 performed close to our campaign, which revealed a relativistic ejection with a reported inclination $\leq45^{\degr}$\, \citep{russell2018}. 
Given that X-ray spectral measurements seem to indicate a highly inclined disk \citep{miller2018,xu2018,sreehari2019}, this suggests the presence of a misalignment between the orbital spin and jet axis, which may cause the jet precession. Rapid changes of jet orientation on the sky, recently seen in V404~Cyg and Cir~X-1 \citep[see e.g][]{coriat2019,miller-jones2019}, support the possibility for such precession. 

{We notice that a combination of the two scenarios is also possible. For instance, the stronger IR variability observed at lower frequencies may indicate the presence of a significant contribution from the jet \citep[][]{veledina2011,malzac2018}. Therefore, the weak QPO signal measured in this dataset, could then be simply be reproduced by assuming a higher $\tau$ and B (i.e.  smaller radii with higher density and magnetic energy density), as the IR band would be  falling in the self-absorption regime. A detailed simulation is however necessary in order to disentangle the contribution from these two components and to better understand the observed differences with respect to GX 339-4.}

\subsection{Cross-correlation analysis}

Our analysis reveals a clear connection between the X-ray and the IR bands. The CCF presents a significant asymmetric anti-correlation at positive lags, without any evidence of positive correlation. Anti-correlations between X-ray and O-IR variability have already been seen in other BH LMXBs. 
One of the the most notable cases is probably the optical/X-ray CCF of Swift~J1753.5$-$0127, which showed a strong anti-correlation at negative lags and a positive correlation at positive lags \citep{durant2008,veledina2017}. 
Such behavior was successfully reproduced by invoking a combination synchrotron self-Compton radiation of the hot flow and reprocessing from the outer disc \citep{veledina2011,veledina2013}.
The CCF observed in MAXI~J1535$-$571, instead, shows an anti-correlation at positive lags and no significant (above noise level) positive correlations.
A roughly similar CCF showing an anti-correlation with a positive lag of $\approx$1-2 s (which included though also a positive correlated peak at $\approx$5 s) has also been observed in Swift~J1753.5$-$0127 during the declining phase of the 2005 outburst \citep[][]{hynes2009,veledina2017} and has been reproduced by the hot flow model, by invoking a second source of X-ray photons \citep[see Fig. 5 in ][]{veledina2017}.
It is interesting to notice that Swift~J1753.5$-$0127, as for our observations, was in the hard-intermediate state when such anti-correlation was observed. 
%

In the Fourier domain, we detect a roughly constant negative phase lags $-\pi/2$ below $\sim$1~Hz.
Given the similar shape of the CCF, the expected phase lags from the hot-inflow model are in good agreement with this result. We notice, however, that similar negative phase lags have also been observed in BH LMXB GX~339$-$4 and have been successfully explained with the internal shocks model \citep{malzac2018}. 
According to this scenario such a behaviour is mainly due to the differential response of the shocks to the input (X-ray) fluctuations \footnote{In the internal shocks model \citep{malzac2014}, the shell velocity is proportional to the X-ray fluctuations. The shocks occur because of the difference in velocity between the shells, which leads to a differential response.}. An impulse response function defined as the derivative of a function is known to give a phase lag of $\pi/2$ \citep[see also][]{jenkins,malzac2004}. Depending on the inclination and on the Lorentz factor of the jet, such an effect can become more or less dominant, changing the phase lag \citep[see Fig. 13 and 14 in][]{malzac2018}.

Always in this context it is not clear why the 0.1~s infrared lag, usually associated to the jet scenario, is not present. From the simulations shown in \citet{malzac2018}, the lag is expected in all tested configurations, and should not depend on the jet Lorentz factor, nor on the inclination. However, in the internal shocks scenario the dependence of the IR variability on the X-rays fluctuation is highly non-linear and difficult to predict without detailed simulations.  It is possible that the QPOs dominating the X-ray PSD for frequencies higher than $\approx$2~Hz prevent the detection of the 0.1~s lag. A similar situation was already found in GX~339$-$4 \citep{malzac2018} where the presence of a low-frequency QPO was thought to pollute the $-\pi/2$ phase lag expected in internal shocks scenario.
Another possibility is that the 0.1~s lag dominates the lag spectrum at higher frequencies than usually observed. \citet{paice2019} showed that -- at least In the case of MAXI~J1820+070 -- indeed the 0.1~s lag can be observed up to at least 5~Hz. 

\subsection{Mid-IR}

We present for the first time the analysis of the correlated variability between simultaneous X-rays, near and mid-IR data. No significant correlation was detected during the first night, while during the second one correlation was found at a $\approx3 \sigma$ level, dropping to below 2 $\sigma$ once the long-term trend of the light curves is removed. Such a behaviour can be explained by the fact that both X-ray and near-IR PDS show a low-frequency break at $\approx$0.03 Hz (30 s timescales).  This means that on the timescales probed by the mid-IR observation, both the near-IR and the X-ray light curves show a small amount of variability, hampering the possibility to detect any correlation.

A detailed discussion of the  mid-IR observations of MAXI~J1535$-$571 carried out during our campaign has been presented in \citet{baglio2018}. The authors reported significant mid-IR variability on timescales of minutes which, given also the strong excess shown in the spectral energy distribution (SED), was interpreted in terms of synchrotron emission from a collimated jet. Contribution from the hot-inflow at these wavelengths seems also be  unlikely when considering the preliminary parameters obtained in the previous section. The same authors, instead, conclude that in the near-IR band there may be potential contribution from both jet and accretion disc \citep{baglio2018,russell2020}. We therefore discuss our results regarding the mid/near-IR connection according to these two possible scenarios:

\textit{Both mid-IR  and near-IR from jet}: Being part of the same physical component, a  tight correlation  should be present between the two bands, with possible delays of the order of few seconds \citep[][]{malzac2018}.  We notice, however, that although we do not find correlation on timescales of tens of seconds, this does not necessarily point against this scenario. As mentioned above, neither X-ray nor near-IR emission display strong variability on timescales comparable with VISIR's time resolution: thus, a low connection is somehow expected in this case. New detailed simulations which include the low-frequency end of the X-ray fluctuation PSD may help to quantitatively test this scenario with this dataset.

\textit{Mid-IR jet, near-IR hot inflow}: On short timescales, the expected correlation strongly depends on the physical parameters of the system, which can affect the responses of the two components. The shape of the CCF may have a complex shape difficult to predict without proper modelling. Nevertheless, if the same input mass-accretion rate fluctuations travel through the hot-inflow (where it emits in the near-IR) and then in the jet (emitting in the mid-IR), a lag of the order of seconds is expected to appear. On longer timescales, instead, the two bands should be correlated, according to the well known strong inflow-outflow connection of these sources \citep[see e.g.][ and references there in]{gallo2018}.

Given that both scenarios can reproduce the observed correlation on longer timescales, a self consistent modelling with both the  components is necessary in order to actually quantify the contribution of hot-inflow and jet in the near-IR. More importantly, the two scenarios  have strongly different predictions when probing (sub-)second timescales, which could not be reached with the current dataset. Thus, new higher time resolution observations will provide crucial physical information.

\section{Conclusion}

In this paper we report the discovery of correlated X-ray/near-IR variability from the BH LMXB MAXI~J1535$-$571, studied with two observations on two consecutive nights in September 2017.  We can summarise our results as follows:

\begin{itemize}
\item Power spectral analysis reveals the presence of complex broadband noise down to sub-second timescales in both  X-ray and near-IR bands. 
No significant differences were found in the PSDs during the two consecutive nights of our campaign. 
A QPO at $\approx 2.1$~Hz is found both in X-rays and in the near-IR. 
The face-value significance of the near-IR QPO is only 2$\sigma$, but the fact that the measured lags at the QPO frequencies have relatively small uncertainties demonstrates the QPO is present in the near-IR light curve.  
This is the first unambiguous detection of a near-IR and X-ray QPO at the same frequency. 
The high QPO frequency and nearly zero phase lag ({with $\pi$/6 radians 3 $\sigma$ upper limit}) between the bands puts tight constraints on the models.
We discuss the origin of the near-IR QPOs in terms of two specific scenarios: Lense-Thirring precession of the hot accretion flow \citep{veledina_poutanen_ingram_2013} or the simultaneous precession of the jet and the inflow \citep{malzac2018}. Both scenarios provide constraints that seem to confirm previous indications of a misalignment between the disk and the jet.

\item The cross-correlation function shows a puzzling asymmetric anti-correlation at positive lags. 
Such feature corresponds to nearly constant phase lags $\approx-\pi/2$ (X-rays lag behind the near-IR).
Similar behaviour was observed in GX~339$-$4 and explained in that case in the context of the internal shocks model  \citep{malzac2018,vincentelli2019}, and during the hard-intermediate state of Swift~1753.5$-$0127, reproduced by the hot inflow model involving two X-ray continua \citep{veledina2017}.

\item We performed the first correlated analysis of the variable mid-IR (4.85--11.88 $\mu$m) and near-IR (2.2 $\mu$m) emission from a BH LMXB. No significant correlation is found during the first night, because of the lack of variability of the near-IR lightcuve on timescales probed by the VISIR observation. No significant correlation is found on the timescales up to ~400 s. The flux-flux correlation diagram for the second night revealed instead a 3$\sigma$ level correlation, likely associated with a clearly visible long-term trend in the mid-IR light curve. The trends can be associated with oscillations on timescales much longer than the corresponding segment of the light curve. In our case, from the trend visible in the lower panel of Fig. \ref{fig:all_lc}, we can only say that the oscillation has a characteristic timescale $\gg$1000s.
 
\end{itemize}

Fast multi-wavelength variability is revealing to be one of the most powerful tools to study the innermost regions around accreting stellar-mass BHs, allowing us to put constraints on the accretion geometry of these systems.
The analysis of an increasing number of datasets, together with the development of theoretical models, are starting to shed light also on the physical processes which take place inside the jet. 
In particular, the results found in our analysis will provide a solid test for the state-of-the-art models.
New high time-resolution multiwavelength observations, especially at longer wavelengths \citep[e.g. with {\it JWST};][]{gardner2006}, will help to address many remaining open questions on these objects.

\section*{DATA AVAILABILITY STATEMENT}

X-ray data is accessible from the XMM-Newton online archive\footnote{\url{http://nxsa.esac.esa.int/nxsa-web/search}}. Data from HAWK-I and VISIR instead are publicly available on the ESO online archive \footnote{\url{http://archive.eso.org/eso/eso_archive_main.html}}

\section*{Acknowledgements}

The authors are extremely grateful to the VLT and XMM-Newton astronomers, who allowed us to successfully collect strictly simultaneous data from 3 different telescopes. {The authors would like to thank the referee for the useful comments which significantly improved the paper.}The discussion of this paper benefited from the meeting 'Looking at the disc-jet coupling from different angles' held at the International Space Science Institute in Bern, Switzerland. FMV acknowledges support from STFC under grant ST/R000638/1.   This work was supported by the Programme National des Hautes Energies of CNRS/INSU with INP and IN2P3, co-funded by CEA and CNES. AV acknowledges the Academy of Finland grants 309308 and 321722. 

\bibliographystyle{mnras}
\bibliography{bib} 
\bsp	
\label{lastpage}

\end{document}